\documentclass{ifacconf}

\usepackage{cite}
\usepackage{graphicx}      
\usepackage{natbib}        
\usepackage{algorithm}
\usepackage{algpseudocode}

\usepackage{amsmath,amssymb,amsfonts}
\usepackage{physics}
\usepackage{graphicx}
\usepackage{textcomp}
\usepackage{xcolor}
\usepackage{caption}
\usepackage{subcaption}
\usepackage{bm}
\usepackage{lipsum}
\usepackage{comment}
\usepackage[super]{nth}

\begin{document}
\begin{frontmatter}

\title{Enhanced Flight Envelope Protection: A Novel Reinforcement Learning Approach}


\author[First]{Akin Catak, Ege C. Altunkaya} 
\author[Second]{Mustafa Demir}
\author[Third]{Emre Koyuncu} 
\author[Fourth]{Ibrahim Ozkol} 

\thanks[footnoteinfo]{This work has been submitted to IFAC for possible publication.}

\address[First]{Aviation Institute, Aerospace Research Center, 
    Istanbul Technical University, Istanbul, Türkiye (e-mail: catak15@itu.edu.tr, altunkaya16@itu.edu.tr)}
\address[Second]{Turkish Aerospace, Air Vehicle Technologies Research Center, Istanbul, Türkiye, (e-mail: mustafa.demir@tai.com.tr)}
\address[Third]{Department of Aeronautics Engineering, 
   Istanbul Technical University, Istanbul, Türkiye, (e-mail: emre.koyuncu@itu.edu.tr)}
\address[Fourth]{Department of Aeronautics Engineering, 
   Istanbul Technical University, Istanbul, Türkiye, (e-mail: ozkol@itu.edu.tr)}

\begin{abstract}                

This paper introduces a flight envelope protection algorithm on a longitudinal axis that leverages reinforcement learning (RL). By considering limits on variables such as angle of attack, load factor, and pitch rate, the algorithm counteracts excessive pilot or control commands with restoring actions. Unlike traditional methods requiring manual tuning, RL facilitates the approximation of complex functions within the trained model, streamlining the design process. This study demonstrates the promising results of RL in enhancing flight envelope protection, offering a novel and easy-to-scale method for safety-ensured flight.

\end{abstract}


\end{frontmatter}

\section{Introduction}

Ensuring the safety of an aircraft's operational capabilities has always been an irreplaceable priority. Regardless of the class of the aircraft, stability and safety must be proven at every single operation point within the flight envelope; for this purpose, an intensive effort is made throughout the design. Nevertheless, the control system design undertakes a significant portion of this workload, particularly in the design of fighter jets, which have been designed as aerodynamically unstable for over four decades. Within this context, the design of flight envelope protection algorithms is one of the spearhead requirements in terms of ensuring both safety and stability. 

The principal purpose of flight envelope protection algorithms is averting the aircraft from the deviation of the predetermined safe flight envelope, thereby mitigating the risk of entering an upset condition. Generally, the objective is achieved by limiting the angle of attack, sideslip angle, load factor, angular rates, bank angle, and true airspeed. The distinctions between the algorithms arise at the point of how to limit the aircraft within the safe boundaries and how to recover the aircraft if a violation occurs.

\subsection{State-of-Art}

Various approaches to flight envelope protection are available, including adaptive flight envelope protection algorithms \cite{fep2}, haptic feedback design \cite{fep3}, reachability analysis \cite{fep4}, command and reference governor techniques \cite{fep5, fep6, fep8}, reference command regeneration \cite{fep7}, control barrier functions \cite{fepEnBirinci}, and traditional flight envelope protection algorithms \cite{fep1, fep9}. Specifically, the study in \cite{fep4} introduces an online method for determining the flight envelope using reachability analysis and safeguarding it with optimal control inputs under icing conditions. Given that icing severely degrades the aerodynamic performance of the aircraft, the reliability of a conventional flight envelope diminishes; thus, it is imperative to assess external factors to avoid departing from the safe flight envelope. The proposed method employs a protection strategy that requires meticulous design and validation to ensure effectiveness under all possible scenarios. Additionally, a command governor approach is suggested in \cite{fep5} to limit maneuvers based on load factor and angle of attack. This protection design is tested in a hardware-in-the-loop environment to demonstrate its efficacy; however, a significant limitation of this method is the need for scheduling the weights in the constructed objective function. This scheduling issue is similarly noted in \cite{fep8}, where there is a necessity for the scheduling of the command governor matrices.

\subsection{Problem Statement}

Even though there is a diverse variety of flight envelope protection algorithms in the literature, their industrial design relies on classical methods. However, the design of classical flight envelope protection algorithms requires careful, experienced, and wise evaluation to certify safety within the entire flight envelope of the aircraft. Otherwise, improper design may result in high pilot workload, unacceptable flying quality, and even instability. As a result, the conventional design approach can be regarded as a tedious and serious workload. 

In this regard, the study aims to address both safety and pilot workload issues while proposing a fast and reliable solution approach. Therefore, a novel flight envelope protection algorithm using reinforcement learning is presented to address the aforementioned issues effectively, and as the primary step of the research, a flight envelope protection algorithm just for the angle of attack and load factor is developed. As a consequence of the research, a care-free, i.e., less pilot workload, and safety-ensured flight envelope protection system is expected.

\subsection{Contributions}


The contributions of this study are itemized as follows;

\begin{itemize}
    \item A novel flight envelope protection algorithm leveraging reinforcement learning is introduced.
    \item The issues encountered throughout the design of conventional flight envelope protection algorithms are mitigated with the proposed method.
    \item The performance deterioration of traditional flight envelope protection algorithms during extreme maneuvers, where coupling effects and nonlinearities arise remarkably, is mitigated with the proposed method.
\end{itemize}

Consequently, the efficacy of the proposed method is assessed under diverse circumstances.

\section{Preliminaries}

The fundamental requirements prior to constructing the principal methodology are briefly presented in this section. 

\subsection{Nonlinear Flight Dynamics}
\label{dynamicsIntro}

The minimum needs for developing a nonlinear flight dynamics model for F-16 is discussed under two basic headlines: the equations of motion as well as aerodynamics and actuators.

\subsubsection{Equations of Motion}
\label{eqofMot}

The nonlinear rigid body dynamics are given in a compact form in Eq.~\eqref{translationalDynamics} and Eq.~\eqref{rotationalDynamics}. 

\begin{equation}
\label{translationalDynamics}
    \bm{\dot{V}} = m^{-1}[\bm{F} -  \bm{\omega} \times m \bm{V}]
\end{equation}  

\begin{equation}
\label{rotationalDynamics}
    \bm{\dot{\omega}} = J^{-1}[\bm{M} -  \bm{\omega} \times J \bm{\omega}]
\end{equation}  
where $\bm{V} \in \mathbb{R}^{3 \times 1}$ is the body velocity vector, $\bm{\omega} \in \mathbb{R}^{3 \times 1}$ is the angular rate vector, $\bm{F} \in \mathbb{R}^{3 \times 1}$ is the total body force vector, $\bm{M} \in \mathbb{R}^{3 \times 1}$ is the total moment vector, $J \in \mathbb{R}^{3 \times 3}$ is the inertia tensor, and $m$ is the mass of the aircraft. Additionally, the rotational kinematics are given by Eq.~\eqref{rotationalKinematics}.

\begin{equation}
\label{rotationalKinematics}
    \bm{\dot{\Omega}} = 
\begin{bmatrix}
1 & \sin\phi\tan\theta & \cos\phi\tan\theta \\
0 & \cos\phi & -\sin\phi \\
0 & \sin\phi \sec\theta & \cos\phi \sec\theta
\end{bmatrix} \bm{\omega}
\end{equation}  
where $\bm{\Omega}$ denotes the Euler angles.

\subsubsection{Aerodynamics and Actuators}
\label{aeroModel}

The aerodynamic modeling consists of the scheduled wind-tunnel data, provided in \cite{f16Data}, and their corresponding formulation to obtain the force ($C_x, C_y, C_z$) and moment coefficients ($C_l, C_m, C_n$). After calculating the coefficients, the necessary dimensionalization is processed using the current states and geometric properties of the aircraft, i.e. dynamic pressure, wing area, span, and mean aerodynamic chord. 

In addition, each control surface has a unique actuator modeling, i.e. a first-order model with a time constant $0.0495s$. Furthermore, the rate saturation values are $60^\circ/s$, $80^\circ/s$, and $120^\circ/s$, whereas the position saturation values are $\pm 25^\circ$, $\pm 21.5^\circ$, and $\pm 30^\circ$ for the horizontal tail, aileron, and rudder respectively.

\subsection{Flight Control Laws}
The control augmentation system consists of a single-loop angular rate control law using the incremental nonlinear dynamic inversion (INDI). Prior to deriving the control law, the control-affine form of the Euler's equations of motion is given by Eq.~\eqref{rotationalDynamics2}.

\begin{equation}
\label{rotationalDynamics2}
    \bm{\dot{\omega}} 
    = 
    -J^{-1} (\bm{\omega} \times J \bm{\omega})
    + 
    \underbrace{J^{-1}\bar{q}_\infty S 
    \begin{bmatrix}
      b &  &  \\
      & \bar{c} &   \\
      &   & b \\
    \end{bmatrix} \Phi}_{\substack{\bm{g}(\bm{x})}} 
    \underbrace{\bm{\delta}}_{\substack{\bm{u}}} 
\end{equation} 
where $\Bar{q}_\infty$, $S$, $b$, and $\Bar{c}$ are the dynamic pressure, wing area, wing span, and mean aerodynamic chord, respectively. Additionally, $\Phi \in \mathbb{R}^{3 \times n}$ denotes the control effectivity matrix, with $n$ is the number of control surfaces. Finally, the control law for the angular rates is presented in Eq.~\eqref{INDIlaw}.

\begin{equation}
\label{INDIlaw}
    \bm{u} = \bm{g}(\bm{x}_0)^{-1}[\bm{\Dot{\omega}}_c - \bm{\Dot{\omega}}_0] + \bm{u}_0 
\end{equation} 
where the subscript $"0"$ denotes the current state and $\bm{\Dot{\omega}}_c \in \mathbb{R}^3$ is the virtual input to be designed. The exact form of the control law is given by Eq.~\eqref{INDIlaw2}. 

\begin{equation}
\label{INDIlaw2}
    \bm{\delta} = \Bigg\{J^{-1}\bar{q}_\infty S 
    \begin{bmatrix}
      b &  &  \\
      & \bar{c} &   \\
      &   & b \\
    \end{bmatrix} \Phi \Bigg\}^{-1}[\bm{\Dot{\omega}}_c - \bm{\Dot{\omega}}_0] + \bm{\delta}_0 
\end{equation} 
where $\bm{\delta} \in \mathbb{R}^3$ signifies the control surface deflections, i.e. aileron, horizontal tail, and rudder, respectively. Furthermore, the virtual input $\bm{\Dot{\omega}}_c$ is provided by Eq.~\eqref{eq:fca13}.

\begin{equation} \label{eq:fca13}
\bm{\Dot{\omega}}_c
= 
\begin{bmatrix}
K_p & & \\
 & K_q & \\
 & & K_r \\
\end{bmatrix} 
\begin{bmatrix}
p_{c} - p \\
q_{c} - q \\
r_{c} - r \\
\end{bmatrix} 
\end{equation}
where $K_p$, $K_q$, and $K_r$ represent the gains for the roll, pitch, and yaw channels, respectively. 

Consequently, the aforementioned relations complete the flight control law for the angular rates.

\subsection{Classical Flight Envelope Protection}

The classical protection algorithm for the angle of attack and load factor is developed based on a simple rationale: applying a counter-action as a restorative measure to prevent FEP violations.

\subsubsection{Angle of Attack and Load Factor Protection}

\begin{figure}[hbt!]
\centering
\includegraphics[width=3.4in]{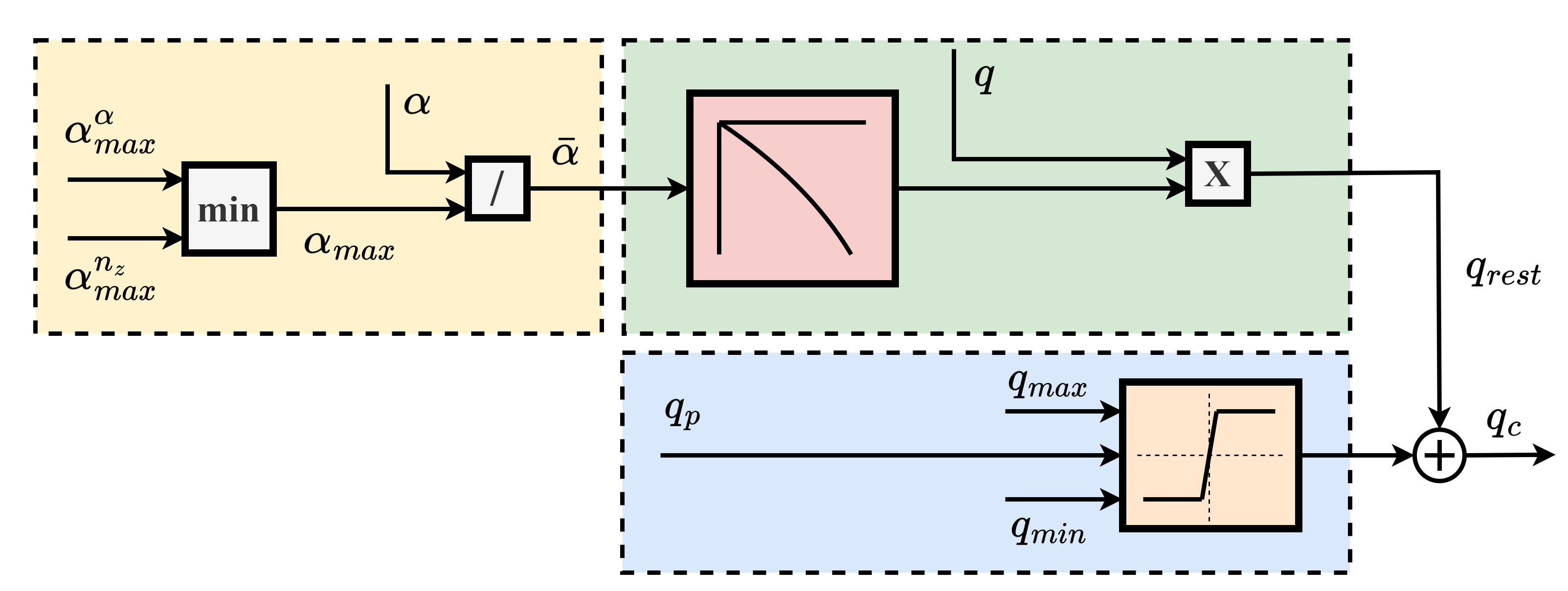}
\caption{The classical angle of attack and load factor protection scheme.}
\label{fig:longProt}
\end{figure}

The excess angle of attack and load factor boundaries can be prevented through the implementation of a 'restorative' pitch rate command. Before reaching the predefined limits of the angle of attack and load factor, the pilot commands are gradually diminished to avert the aircraft's entry into an upset condition. Additionally, both angle of attack and load factor protection must be viewed comprehensively to produce a singular and cohesive restorative pitch response; hence, the load factor boundary is also translated into the angle of attack as represented by Eq.~\eqref{loadFactor2Alpha}.

\begin{equation}
\label{loadFactor2Alpha}
    \alpha_{max}^{n_z} = \dfrac{W n_{z_{max}}}{\Bar{q}_\infty S C_{z_\alpha}}
\end{equation}
where $\alpha_{max}^{n_z}$ denotes the angle of attack equivalent to the maximum allowable load factor $n_{z_{max}}$, $W$ is the weight of the aircraft, $C_{z_\alpha}$ is the force coefficient in z-direction derivative with the angle of attack. The general framework is depicted in Fig.~\ref{fig:longProt}.

Consequently, the classical flight envelope protection for the angle of attack and load factor is completed.

\subsection{Reinforcement Learning}

Reinforcement Learning (RL) is a subfield of Machine Learning in which the problem is represented as a Markov decision Process (MDP) (\cite{van2012reinforcement}), and the actions taken by the agent are evaluated utilizing Belman optimality conditions. The objective of the RL is to maximize the cumulative reward by finding the optimal policy of the agent while dynamically interacting with the environment. An agent can be described as anything that learns and makes decisions while interacting with the environment. The environment is what the agent interacts with, basically everything except the agent.  In RL, learning is achieved in an iterative way by evaluating the sequence of actions and eventually maximizing the cumulative rewards by updating the policy such that in each state, the policy selects the near-optimal actions as training progresses. 

In MDP every step can be represented with the tuple $<s,a,p,r>$ consisting of environments state ($s\in \mathcal{S}$), action ($a \in \mathcal{A}$), transition function  $p$ ($0\leq p \leq 1 $ ) between state $s$ and state $s'$, and reward $r$ for taking the action $a$ that takes from state $s$ to state $s'$. Another important concept is the policy $\pi(s|a)$, which can be thought of as a strategy. The policy determines the probability of each action at each stage. The objective of Reinforcement learning is to find an optimal policy $\pi^{*}(s|a)$ in which the cumulative reward is maximized. The Value function is the combination of immediate Reward and future rewards. Thus, in other words, the objective can be thought of as finding an optimal policy that maximizes the expected value function.

\subsection{Actor Critic - Deep Deterministic Policy Gradient}

Actor-critic methods are the natural extension of the idea of reinforcement-comparison methods to temporal difference learning and to the full reinforcement-learning problem (\cite{konda1999actor}). It combines the benefits of both value-based and policy-based RL methods. Typically, the critic is a state-value function. After each action selection, the critic evaluates the new state to determine whether things have gone better or worse than expected. Deep Deterministic Policy Gradient is one of the actor-critic algorithms that is used for continuous actions (\cite{lillicrap2015continuous,duan2016benchmarking}). It approximates actor and critic using deep neural networks and employs deterministic policy gradient for learning.

The action determined by the actor with the policy parameters $(\theta^{\pi})$ is given in Eq. \eqref{eq:action_taken_by_actor} with the added stochastic noise $\mathcal{N}_t$
\begin{equation}\label{eq:action_taken_by_actor}
     a_t = \pi(s_t|\theta^{\pi}) + \mathcal{N}_t 
\end{equation}

The critic network is responsible for evaluating the Q-value for the state-action pairs. Critic calculates the temporal difference(TD) error, $\delta_t$, using the immediate reward ($r_t$), current Q-Value, and target models Q-value $Q'$ as given in Eq.~\eqref{eq:TD_error}.

\begin{equation}\label{eq:TD_error}
    \delta_t = r_t + \gamma Q'(s_{t+1}, \pi'(s_{t+1}|\theta^{\pi'})|\theta^{Q'}) - Q(s_t, a_t|\theta^{Q})
\end{equation}
where the $Q'$ and $\pi'$ are target models with their parameters $\theta^{Q'}$ and $\theta^{\pi'}$, and $\gamma$ being discount factor, basically determining the importance of the future states.

Critic parameters $\theta^{\pi}$ are updated by minimizing the loss across all samples as in Eq.~\eqref{eq:Loss_temporal_difference}.

\begin{equation}\label{eq:Loss_temporal_difference}
L=\frac{1}{2 M} \sum_{i=1}^M\left(\delta_t\right)^2
\end{equation}
where $M$ is the mini-batch size.

The actor parameters $\theta^{\pi}$ are updated using the gradient given in Eq. \eqref{eq:actor_parameter_Gradient}. To maximize expected discounted reward 
\begin{equation}\label{eq:actor_parameter_Gradient}
    \nabla_{\theta^{\pi}} J \approx \frac{1}{2 M} \sum_{i=1}^M\left( \nabla_{a} Q(s,a|\theta^{Q})|_{a=\pi(s|\theta^{\pi})} \nabla_{\theta^{\pi}} \pi(s|\theta^{\pi}) \right)
\end{equation} 
where $\nabla_{a} Q(s,a|\theta^{Q})$ is the gradient of the critic output with respect to action ${a=\pi(s|\theta^{\pi})}$ generated by actor network and $\nabla_{\theta^{\pi}} \pi(s|\theta^{\pi})$ is the gradient of the actor network's output with respect to the actor parameters $\theta^{\pi}$.

The target networks are updated with smoothing factor $\tau$ as given in Eq.~\eqref{eq:target_model_update}.

\begin{align}\label{eq:target_model_update}
       &\theta^{Q'} \leftarrow \tau \theta^{Q} + (1 - \tau) \theta^{Q'} \\
     &\theta^{\pi'} \leftarrow \tau \theta^{\pi} + (1 - \tau) \theta^{\pi'}
\end{align}
where small $\tau= 0.001$ enables smooth updates for target networks.

\section{Methodology and Reward Structure}

\begin{figure}[hbt!]
\centering
\includegraphics[width=3.4in]{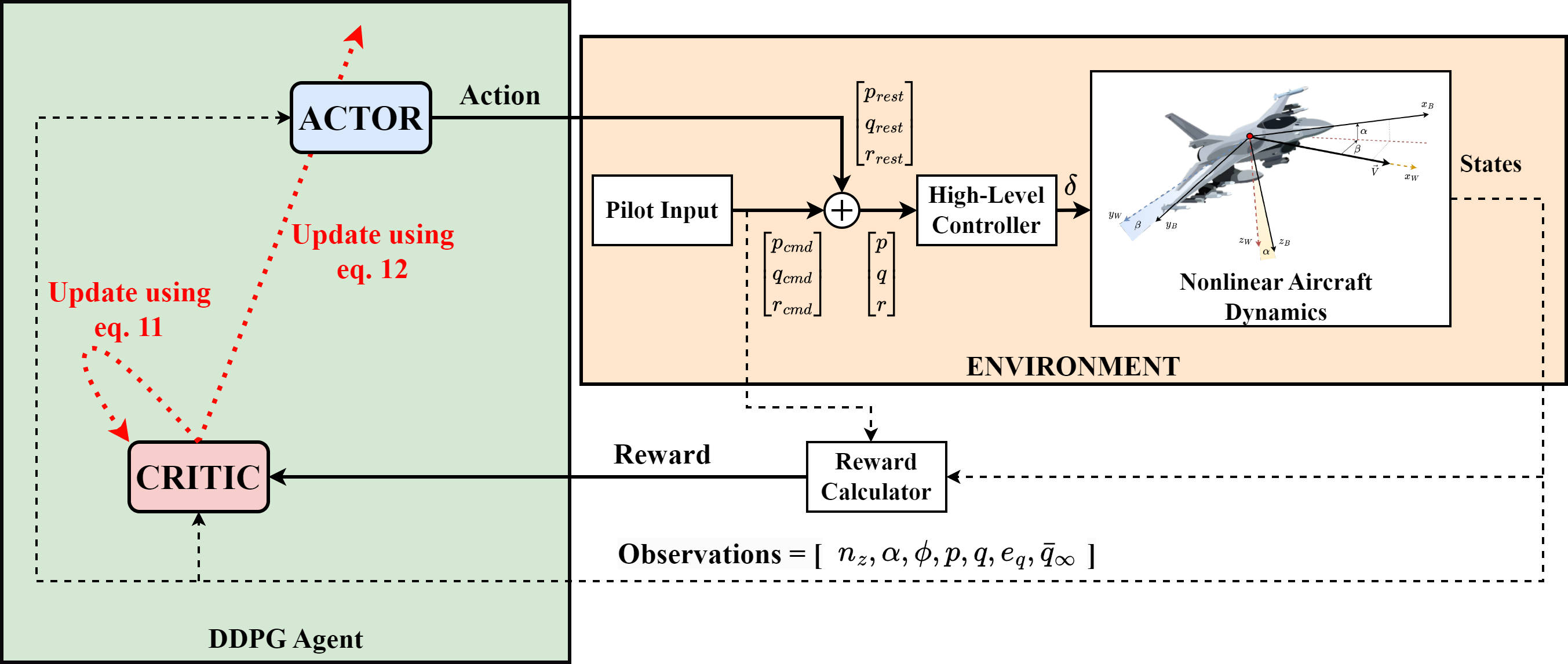}
\caption{Actor-critic scheme.}
\label{Actor Critic Scheme}
\end{figure}

The proposed method focuses on developing a flight envelope protection scheme given in Fig. \ref{Actor Critic Scheme} using RL. In this method, the environment consists of the nonlinear F-16 model and the flight control law. This environment enables capturing the complex flight dynamics and control challenges. Since we have continuous actions, the Deep Deterministic Policy Gradient (DDPG) algorithm, an off-policy actor-critic method that is also sample efficient, is employed. The action of the proposed envelope protection scheme in the longitudinal axis is the restorative action $q_{rest}$ within the range $-20^\circ/s \leq q_{rest} \leq 30^\circ/s$. 

For the proposed method, observation space should contain the relevant states of the environment, in this case, the nonlinear F-16 model. 

\begin{equation}
    S = [n_z, \alpha, \phi, p, q, e_q, \Bar{q}_\infty]
\end{equation}
where states $n_z, \alpha$ and $q$ that need to be protected are included as well as $\phi$ and $p$ due to coupling affects. The tracking error, $e_q = q_{cmd}-q$, is included to give information about tracking error and dynamic pressure $\Bar{q}_\infty$ is also included since its substantial effects over the dynamics. In training, the aircraft starts from level flight trim condition with 0.6 Mach speed and 500-meter altitude. Each episode is 10 seconds long and has constant pilot input on the pitch channel determined randomly at the beginning of each episode. The training is stopped when the average reward of 150 episodes reaches the desired stopping value.

 The reward used in the proposed method is in the form of feedback taken from the environment, which evaluates how good or bad the current policy is given the desired objectives. For our problem, the Reward function is in the form of Eq.~\eqref{eq:reward eq}.

\begin{equation} \label{eq:reward eq}
R = r_s + W_t  r_t + W_a  r_a + W_n  r_n + W_q  r_q + r_p
\end{equation}
where $r$s represent individual rewards and  $w$s are corresponding weights. $r_s$ is the positive constant reward obtained at each step supporting the agent to survive. $r_t$ is the tracking cost added for increasing the track performance. $r_a$ is the cost of the angle of attack excess, which will be defined later. Similar to $r_t$, $r_n$ and $r_q$ are the costs for exceeding load factor and pitch rate limits respectively. The last component of $R$ is the $r_p$, which is the so-called isdone reward penalizing the worst policies.

Reward related to survival is  $r_s = 0.1$. Since it is a constant it does not have a weight associated with it. The cost related to tracking performance is given in Eq. \eqref{eq:trackıng reward}.

\begin{equation} \label{eq:trackıng reward}
r_t=\left(\frac{\left(|q|-\left|q_{c m d}\right|\right)}{\left|q_{c m d}\right|+\epsilon}\right)^2
\end{equation}
where $q$ is the pitch rate, $q_{cmd}$ is commanded pitch rate and $\epsilon$ is a small number for avoiding zero division. 

The cost related to exceeding the angle of attack limit is given in Eq. \eqref{eq:aoa reward}. 

\begin{equation} \label{eq:aoa reward}
r_a=\left\{\begin{array}{cl}
-\left(\frac{\left(|\alpha|-0.9\left|\alpha_{\max }\right|\right)}{0.9\left|\alpha_{\max }\right|}\right)^2 & \text{if } \left|\alpha\right| \geq 0.9\left|\alpha_{\max }\right| \\
0 & \text { otherwise }
\end{array}\right.
\end{equation}
where $\alpha$ is the angle of attack and its limit is $\alpha_{max}$. Multiplication with $0.9$ is added to the reward to prevent the agent from getting close to the limit. 
The cost related to exceeding the load factor limit is given in Eq. \eqref{eq:nz reward}.
\begin{equation}\label{eq:nz reward}
r_n=\left\{\begin{array}{cl}
-\left(\frac{\left(|n_z|-\left|n_{zmax }\right|\right)}{\left|n_{zmax }\right|}\right)^2 & \text{if } \left|n_z\right| \geq \left|n_{zmax }\right| \\
0 & \text { otherwise }
\end{array}\right.
\end{equation}
where $n_z$ is the load factor component and $n_{zmax}$ is the maximum allowable load factor. 
Similarly, the pitch rate cost can be defined as well in Eq. \eqref{eq: q reward}.

\begin{equation}\label{eq: q reward}
r_q=\left\{\begin{array}{cl}
-\left(\frac{\left(|q|-\left|q_{max }\right|\right)}{\left|q_{max }\right|}\right)^2 & \text{if } \left|q\right| \geq \left|q_{max }\right| \\
0 & \text { otherwise }
\end{array}\right.
\end{equation}

The last component of the $R$ is the penalty cost, which is designed to penalize the excess of the envelope for a certain time as given in Eq. \eqref{eq: penalty reward}.

\begin{equation}\label{eq: penalty reward}
r_p=\left\{\begin{array}{ll}
-400 & \text {if } cond_1  \operatorname{holds}\\
-600 & \text {if } cond_2  \operatorname{holds}
\end{array}\right\}
\end{equation}
where the ${cond}_1$ is the first condition indicating at least 2 of the envelope parameter($\alpha, n_z ,q$) exceed the limit for 2 seconds and ${cond}_2$ is the second condition indicating at least one of the envelope parameter($\alpha, n_z ,q$) exceed their corresponding limits by $50\%$. If this cost takes place then the episode is terminated.

\section{Results}


In this work, training was conducted using MATLAB\textsuperscript{\tiny\textregistered}'s Reinforcement Learning Toolbox. The computer on which the training was conducted has an Intel i7 $\nth{5}$ series processor with a clock speed of 2.8 GHz and 16 GB of memory. The noise variance is taken as 0.001, the variance decay rate is 1e-9, the actor learn rate is 1e-5, the critic learn rate is 1e-4, and the discount factor is trained as 0.99. The critic network has observation and action paths with having 2 and 1 hidden layer, respectively. The hidden layer on the action path has 40 neurons, while hidden layers on the observation path have 80 and 40 neurons, respectively. Those layers are added together to obtain the Q value. The actor network has one hidden layer with 40 neurons.

The performance of the proposed envelope protection method is examined under three different pilot commands as tabulated in Table~\ref{tab:scenario}. 

\begin{table}[hbt!]
    \centering
    \caption{Simulation scenarios}
    \begin{tabular}{cccc}
        & \multicolumn{3}{c}{Command} \\
        \cline{2-4}
        & $p, ^\circ/s$ & $q, ^\circ/s$ & $r, ^\circ/s$ \\
       \hline
       Scenario-$\#1$ & 0 & 25 & 0 \\
       Scenario-$\#2$ & 0 & -10 & 0 \\
       Scenario-$\#3$ & 60 & $q(t)$ & 0 \\
       \hline
    \end{tabular}
    \label{tab:scenario}
\end{table}

\begin{figure}[hbt!]
\centering
\includegraphics[width=3.4in]{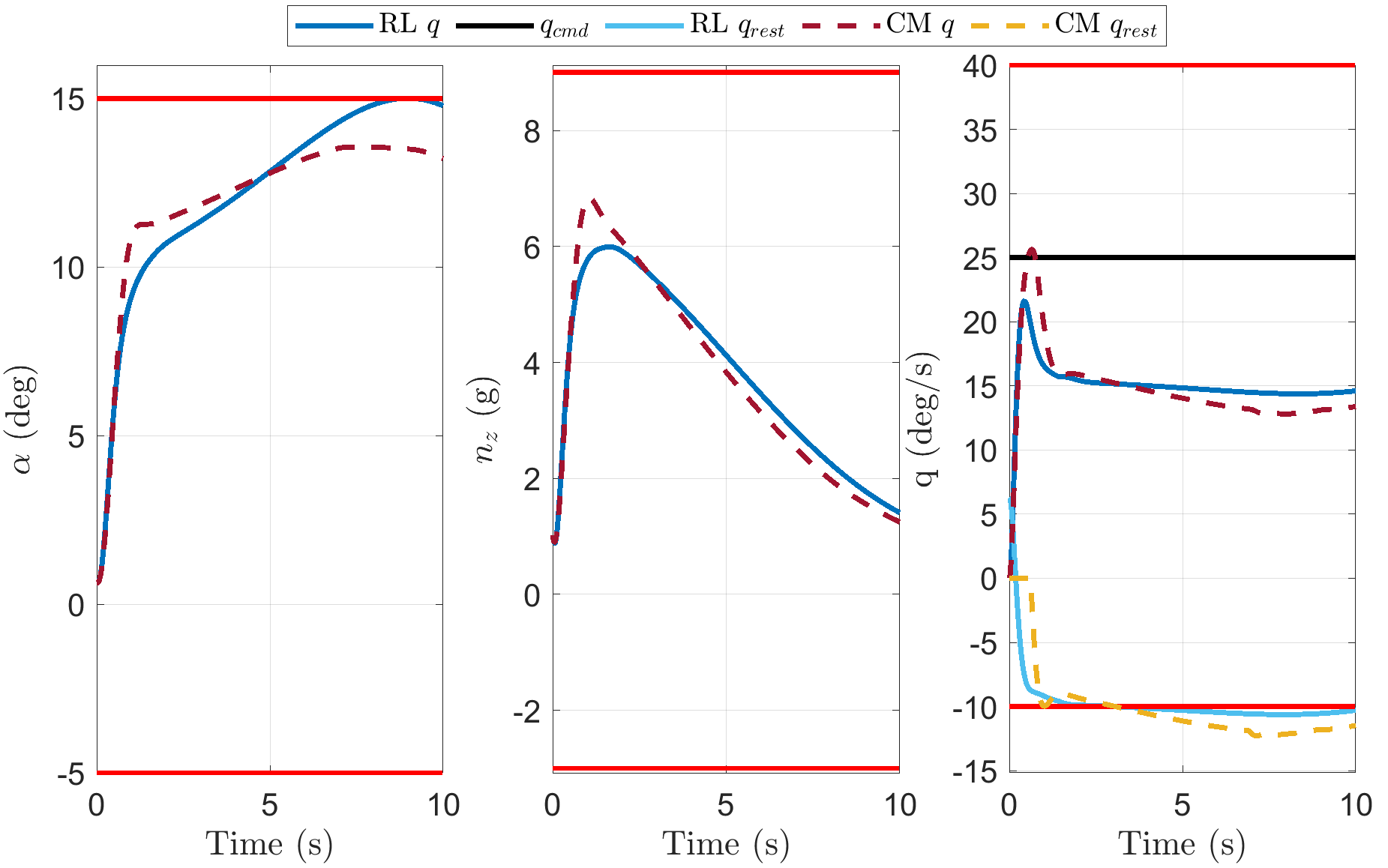}
\caption{The comparison of envelope protection algorithms: scenario-$\#1$.}
\label{fig:25 q_cmd}
\end{figure}

All scenarios include the aforementioned classical envelope protection algorithm as a benchmark to compare. The first scenario includes $q_{cmd} = 25^\circ/s$, which is the maximum pilot command used in the training, and it is evaluated to show the performance of the proposed method in the maximum pilot command, as depicted in Fig.~\ref{fig:25 q_cmd}. 

\begin{figure}[hbt!]
\centering
\includegraphics[width=3.4in]{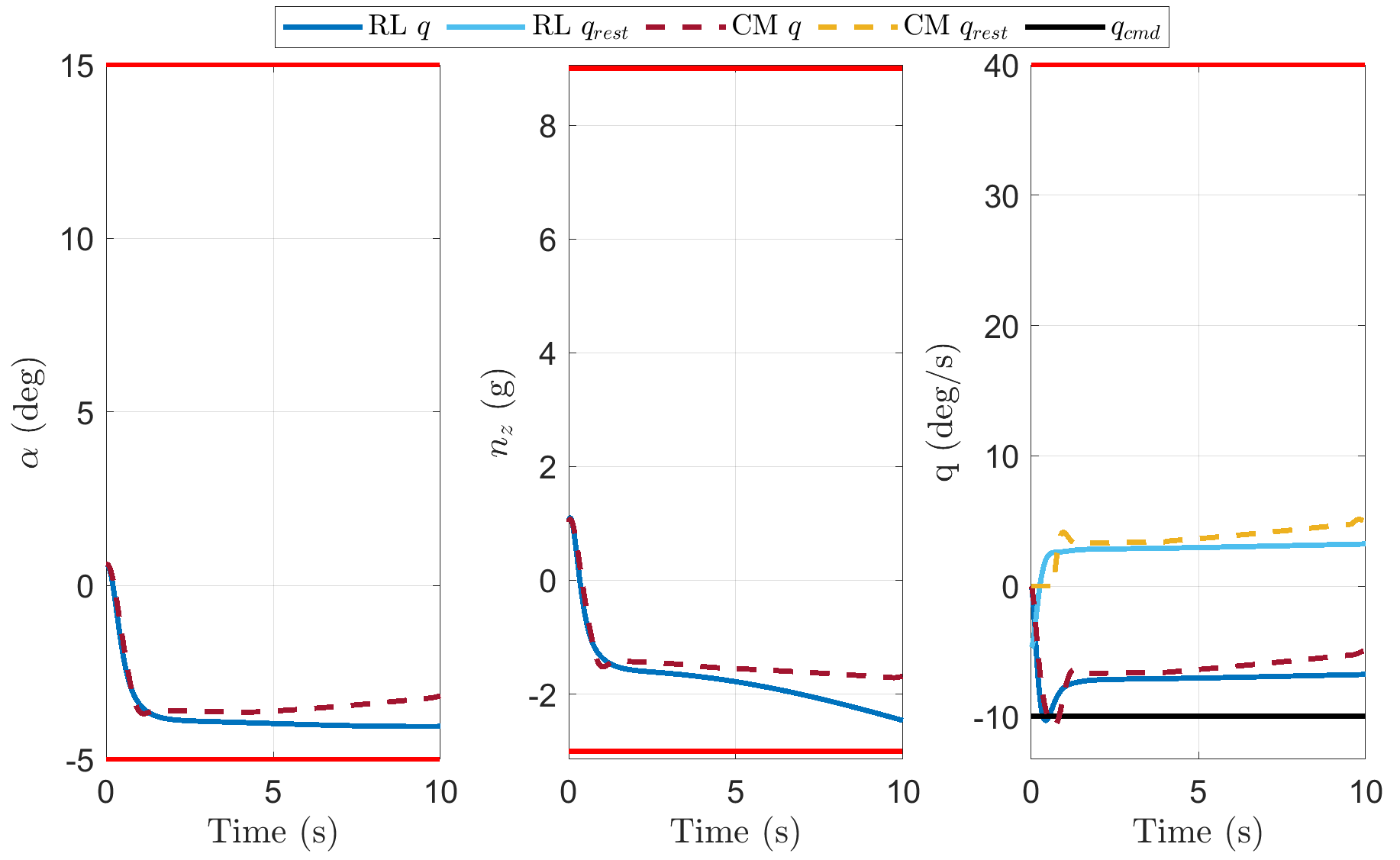}
\caption{The comparison of envelope protection algorithms: scenario-$\#2$.}
\label{fig:-10 q_cmd}
\end{figure}

The red lines indicate the limits for each parameter to be protected. The abbreviation RL in the legend denotes the proposed method, whereas CM stands for the classical approach for the sake of comparison in terms of flight envelope protection performances. This scenario is the most challenging because the angle of attack protection and command tracking performance metrics are conflicting. While the tracking performance is better in the initial phase of the command for the CM, the tracking performance is better for the rest of the simulation, and the angle of attack limit is used more aggressively in the RL method. Both methods can protect the flight envelope in this scenario, but when performance requirements arise, the results are debatable. The results of the second scenario are illustrated in Fig.~\ref{fig:-10 q_cmd}.

\begin{figure}[hbt!]
\centering
\includegraphics[width=3.4in]{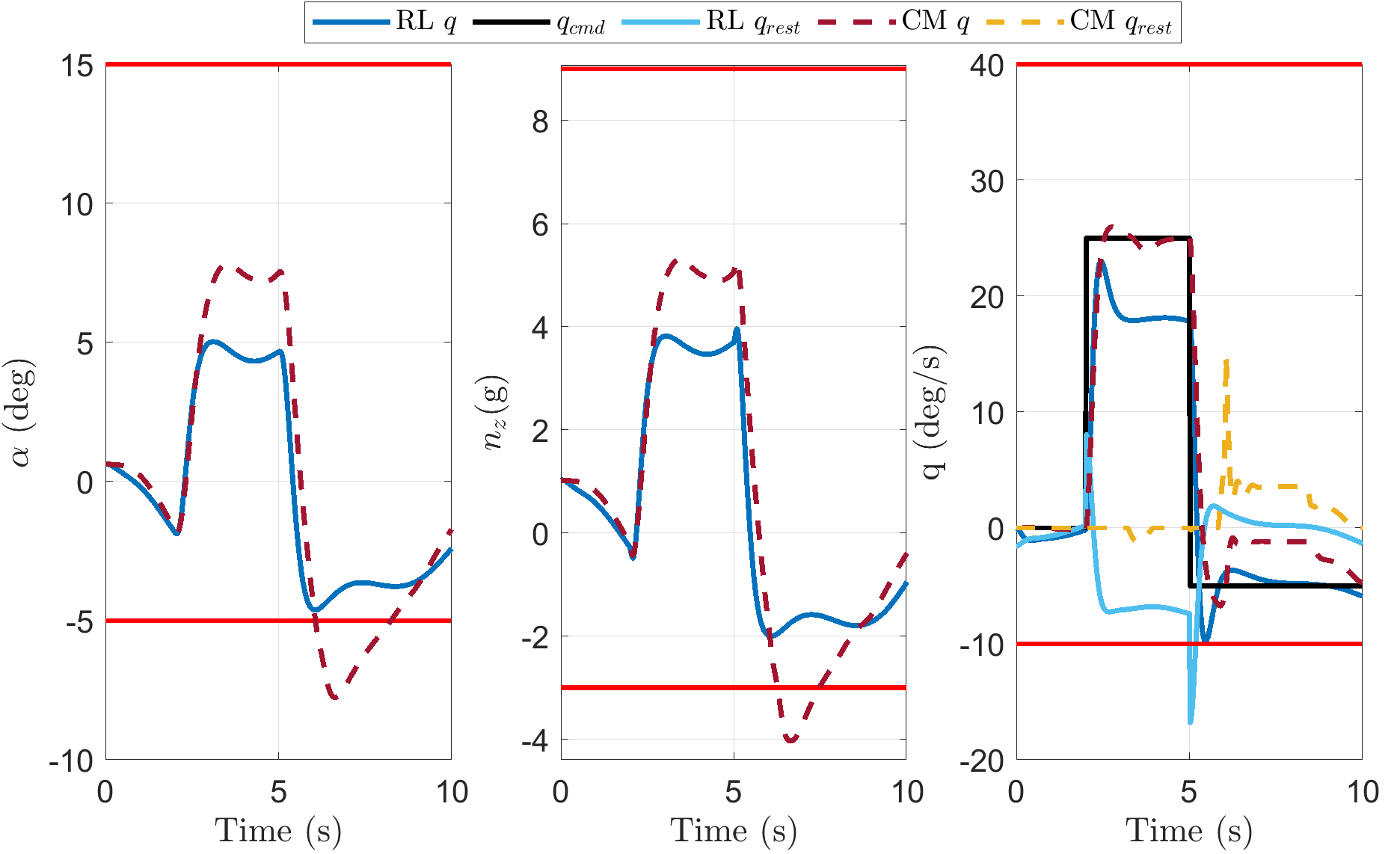}
\caption{The comparison of envelope protection algorithms: scenario-$\#3$.}
\label{fig: coupled command}
\end{figure}

The second scenario is conducted to evaluate the maximum negative pilot command used in the training of the RL, which is actually the negative pitch rate limit of the aircraft. Both methods can protect the envelope but the tracking performance of the RL is slightly better in addition to the better usage of the limits. The third and last scenario is presented in Fig.~\ref{fig: coupled command}.

The third scenario serves as a litmus test, where the proposed RL-based approach is noticeably better in terms of performance and envelope protection. In the initial part of the simulation, involving a positive pitch rate coupled with a roll rate, both methods perform satisfactorily, though CM is slightly better in tracking performance. The crucial part of the simulation is the transition from a positive pitch rate to a negative pitch rate command. This transition is successfully handled by the proposed RL method, while the classical approach fails to protect the envelope. Besides the poor performance in envelope protection, the tracking performance of the classical method is noticeably worse than that of the proposed approach.

\begin{figure}[hbt!]
\centering
\includegraphics[width=3.4in]{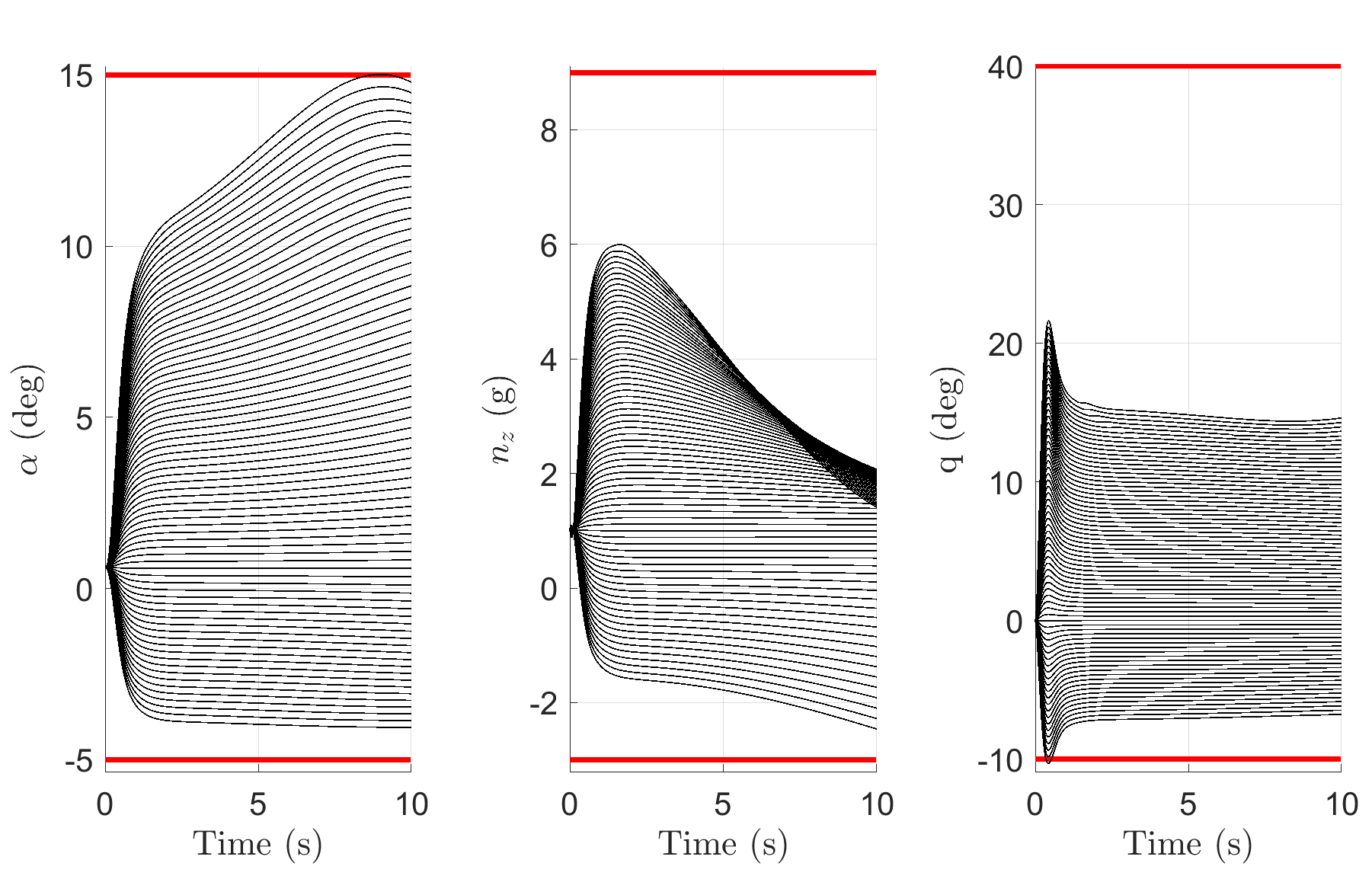}
\caption{Monte Carlo simulation results: pitch-only assessments.}
\label{fig: Monte decoupled}
\end{figure}

Furthermore, Fig.~\ref{fig: Monte decoupled} shows the results for different constant pitch rate commands, demonstrating the proposed method's efficacy in envelope protection under varying circumstances. Each simulation is conducted for different pitch rate commands over an interval of $q_{cmd} = [-10^\circ/s \hspace{0.15cm} 25^\circ/s]$ with an increment of $\Delta q_{cmd} = 0.5^\circ/s$.

The lowest pitch rate command exceeds the negative limit for a short time; however, since the metric for failure is defined as exceeding the limit for at least 2 seconds, this excess is acceptable.

\begin{figure}[hbt!]
\centering
\includegraphics[width=3.4in]{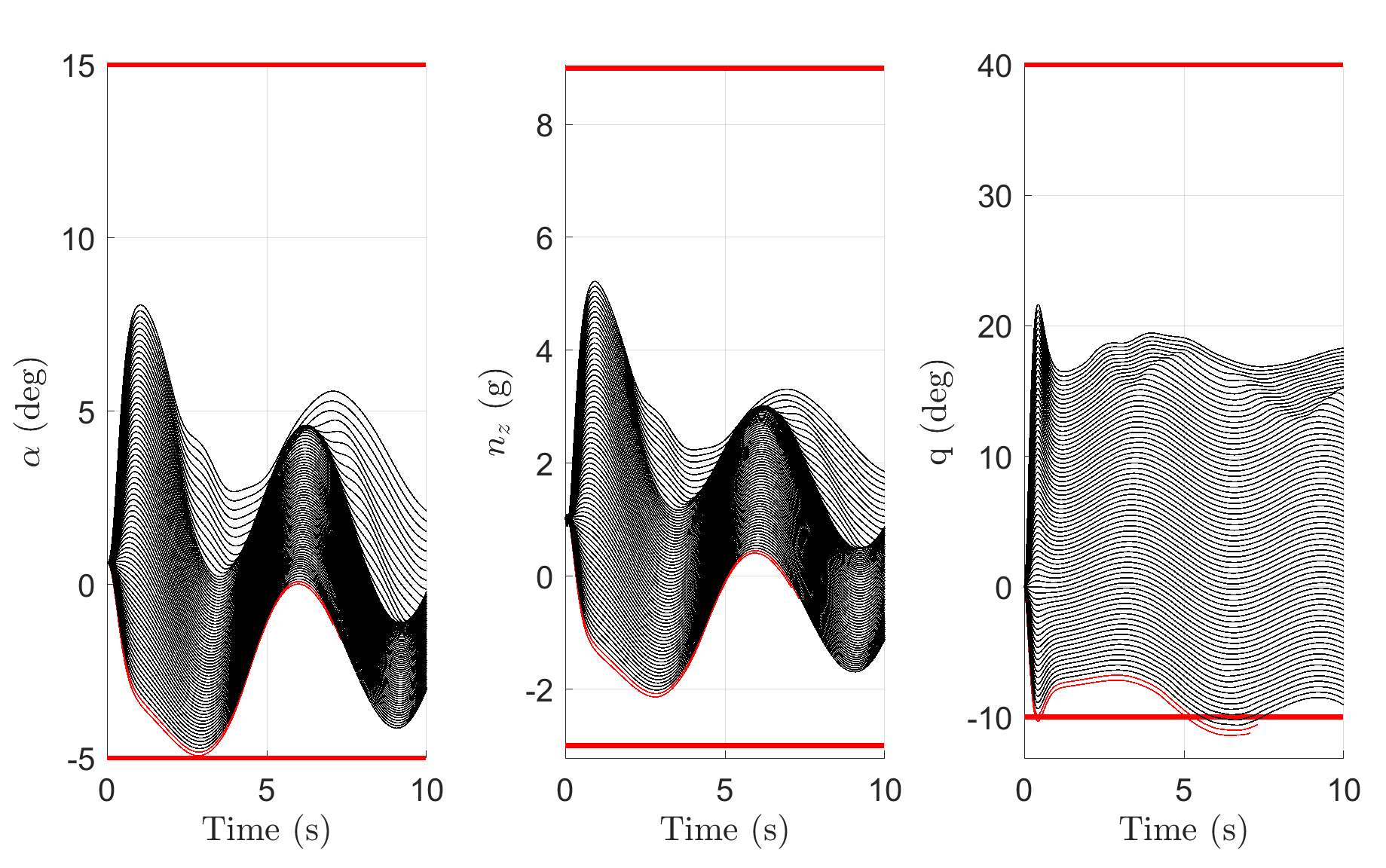}
\caption{Monte Carlo simulation results: pitch-roll coupled assessments.}
\label{fig: Monte coupled}
\end{figure}

Finally, the same analysis is conducted to see the protection performance under pitch-roll coupled maneuver circumstances. In this regard, Fig.~\ref{fig: Monte coupled} is depicted to show the efficacy of the proposed method.

The addition of roll rate command induces oscillation in $\alpha$, $n_z$, and $q$ values. Although the $p$ and $\phi$ are inside the observation space, the weights of the neurons corresponding to these states are relatively small due to zero $p_{cmd}$ throughout the simulated maneuver. It is safe to conclude that the performance of the proposed method is satisfactory in the presence of coupled commands since all the commands are handled except the two red lines.

\section{Conclusion}

In this work, a novel flight envelope protection algorithm for the longitudinal motion of the aircraft is introduced based on reinforcement learning. The algorithm allows the aircraft to conduct considerably aggressive maneuvers without violating the predetermined safe flight envelope. Furthermore, the proposed method surpasses the designed classical flight envelope protection algorithm, especially in pitch-roll coupled maneuvers. As a final assessment, two distinct Monte Carlo simulations are carried out in order to prove the efficacy of the proposed protection algorithm. Impressions from the simulations are not only satisfactory but also show a significant potential for more advanced envelope protection algorithms building on the proposed approach, contributing remarkably to carefree flight. Future plans encompass the studies based on a complete and coherent envelope protection algorithms for both longitudinal and lateral motion to ensure safety, even in the most complex, challenging, and aggressive maneuvers.

\bibliography{ifacconf}            

\begin{thebibliography}{15}
\providecommand{\natexlab}[1]{#1}
\providecommand{\url}[1]{\texttt{#1}}
\providecommand{\urlprefix}{URL }
\expandafter\ifx\csname urlstyle\endcsname\relax
  \providecommand{\doi}[1]{doi:\discretionary{}{}{}#1}\else
  \providecommand{\doi}{doi:\discretionary{}{}{}\begingroup \urlstyle{rm}\Url}\fi

\bibitem[{Altunkaya et~al.(2024)Altunkaya, {\c{C}}atak, Demir, Koyuncu, and {\"O}zkol}]{fepEnBirinci}
Altunkaya, E.C., {\c{C}}atak, A., Demir, M., Koyuncu, E., and {\"O}zkol, {\.I}. (2024).
\newblock Stability and safety assurance of an aircraft: A practical application of control lyapunov and barrier functions.
\newblock \emph{Available at SSRN 4823223}.

\bibitem[{Duan et~al.(2016)Duan, Chen, Houthooft, Schulman, and Abbeel}]{duan2016benchmarking}
Duan, Y., Chen, X., Houthooft, R., Schulman, J., and Abbeel, P. (2016).
\newblock Benchmarking deep reinforcement learning for continuous control.
\newblock In \emph{International conference on machine learning}, 1329--1338. PMLR.

\bibitem[{Konda and Tsitsiklis(1999)}]{konda1999actor}
Konda, V. and Tsitsiklis, J. (1999).
\newblock Actor-critic algorithms.
\newblock \emph{Advances in neural information processing systems}, 12.

\bibitem[{Lillicrap et~al.(2015)Lillicrap, Hunt, Pritzel, Heess, Erez, Tassa, Silver, and Wierstra}]{lillicrap2015continuous}
Lillicrap, T.P., Hunt, J.J., Pritzel, A., Heess, N., Erez, T., Tassa, Y., Silver, D., and Wierstra, D. (2015).
\newblock Continuous control with deep reinforcement learning.
\newblock \emph{arXiv preprint arXiv:1509.02971}.

\bibitem[{Lombaerts et~al.(2017)Lombaerts, Looye, Ellerbroek, and Martin}]{fep2}
Lombaerts, T., Looye, G., Ellerbroek, J., and Martin, M.R.y. (2017).
\newblock Design and piloted simulator evaluation of adaptive safe flight envelope protection algorithm.
\newblock \emph{Journal of Guidance, Control, and Dynamics}, 40(8), 1902--1924.

\bibitem[{Moreira et~al.(2022)Moreira, Gripp, Yoneyama, and Marinho}]{fep1}
Moreira, M.A., Gripp, J.A., Yoneyama, T., and Marinho, C.M. (2022).
\newblock Longitudinal flight control law design with integrated envelope protection.
\newblock \emph{Journal of Guidance, Control, and Dynamics}, 45(9), 1739--1749.

\bibitem[{Nguyen(1979)}]{f16Data}
Nguyen, L.T. (1979).
\newblock \emph{Simulator study of stall/post-stall characteristics of a fighter airplane with relaxed longitudinal static stability}, volume 12854.
\newblock National Aeronautics and Space Administration.

\bibitem[{Nguyen(1980)}]{fep9}
Nguyen, L.T. (1980).
\newblock \emph{Control-system techniques for improved departure/spin resistance for fighter aircraft}, volume 791083.
\newblock National Aeronautics and Space Administration, Scientific and Technical Information Branch.

\bibitem[{Simon et~al.(2017)Simon, H{\"a}rkeg{\aa}rd, and L{\"o}fberg}]{fep5}
Simon, D., H{\"a}rkeg{\aa}rd, O., and L{\"o}fberg, J. (2017).
\newblock Command governor approach to maneuver limiting in fighter aircraft.
\newblock \emph{Journal of Guidance, Control, and Dynamics}, 40(6), 1514--1527.

\bibitem[{Van~Baelen et~al.(2020)Van~Baelen, Ellerbroek, Van~Paassen, and Mulder}]{fep3}
Van~Baelen, D., Ellerbroek, J., Van~Paassen, M., and Mulder, M. (2020).
\newblock Design of a haptic feedback system for flight envelope protection.
\newblock \emph{Journal of Guidance, Control, and Dynamics}, 43(4), 700--714.

\bibitem[{Van~Otterlo and Wiering(2012)}]{van2012reinforcement}
Van~Otterlo, M. and Wiering, M. (2012).
\newblock Reinforcement learning and markov decision processes.
\newblock In \emph{Reinforcement learning: State-of-the-art}, 3--42. Springer.

\bibitem[{Wang et~al.(2022)Wang, Yang, Liu, and Yue}]{fep8}
Wang, L., Yang, K., Liu, H., and Yue, T. (2022).
\newblock Command governor for constrained attitude angle protection of the wing-in-ground effect craft near sea surface.
\newblock \emph{Journal of Aerospace Engineering}, 35(6), 04022080.

\bibitem[{Ye et~al.(2015)Ye, Chen, Wu et~al.}]{fep6}
Ye, H., Chen, M., Wu, Q., et~al. (2015).
\newblock Flight envelope protection control based on reference governor method in high angle of attack maneuver.
\newblock \emph{Mathematical Problems in Engineering}, 2015.

\bibitem[{Yu et~al.(2022)Yu, Zhou, Zhang, Guo, Ye, and Peng}]{fep7}
Yu, X., Zhou, X., Zhang, Y., Guo, L., Ye, S., and Peng, X. (2022).
\newblock Safety control design with flight envelope protection and reference command generation.
\newblock \emph{IEEE Transactions on Aerospace and Electronic Systems}, 58(6), 5835--5848.

\bibitem[{Yu et~al.(2020)Yu, Li, Zhang, Xu, and Dong}]{fep4}
Yu, Z., Li, Y., Zhang, Z., Xu, W., and Dong, Z. (2020).
\newblock Online safe flight envelope protection for icing aircraft based on reachability analysis.
\newblock \emph{International Journal of Aeronautical and Space Sciences}, 1--11.

\end{thebibliography}
\end{document}